\documentclass[preprint]{elsarticle}
\usepackage{graphicx,url,hyperref,microtype,subcaption,graphbox,amsmath}
\usepackage{latexsym,courier,upgreek,xcolor,rotating,ulem}
\usepackage[onehalfspacing]{setspace}
\hypersetup{colorlinks,citecolor=blue,linkcolor=red,urlcolor=green}
\usepackage[displaymath,mathlines]{lineno}
\usepackage[margin=2cm]{geometry}
\modulolinenumbers[5]
\biboptions{numbers,sort&compress}

\begin{document}

\begin{frontmatter}
  
  \title{Modeling crack propagation in heterogeneous materials: Griffith's law, intrinsic crack resistance and avalanches}
  
  \author[add1]{Subhadeep Roy}
  \ead{subhadeeproy03@gmail.com}
  
  \author[add2]{Takahiro Hatano}
  \ead{hatano@ess.sci.osaka-u.ac.jp}
  
  \author[add3]{Purusattam Ray}
  \ead{ray@imsc.res.in}
  
  \address[add1]{PoreLab, Department of Physics, Norwegian University
    of Science and Technology, N-7491 Trondheim, Norway.}
    
  \address[add2]{Department of Earth and Space Science, Osaka University, 560-0043 Osaka, Japan.}
  
  \address[add3]{The Institute of Mathematical Sciences, Taramani, Chennai-600113, India.}

\begin{abstract}
Various kinds of heterogeneity in solids including atomistic discreteness affect the fracture strength as well as the failure dynamics remarkably. Here we study the effects of an initial crack in a discrete model for fracture in heterogeneous materials, known as the fiber bundle model. We find three distinct regimes for fracture dynamics depending on the initial crack size. If the initial crack is smaller than a certain value, it does not affect the rupture dynamics and the critical stress. While for a larger initial crack, the growth of the crack leads to breakdown of the entire system, and the critical stress depends on the crack size in a power-law manner with a nontrivial exponent. The exponent, as well as the limiting crack size, depend on the strength of heterogeneity and the range of stress relaxation in the system.
\end{abstract}

  \date{\today}
  
  
\end{frontmatter}


\makeatletter
\def\ps@pprintTitle{%
 \let\@oddhead\@empty
 \let\@evenhead\@empty
 \def\@oddfoot{}%
 \let\@evenfoot\@oddfoot}
\makeatother

\section{Introduction}
Determination of the strength of materials is a prime objective in material science and applied mechanics. Traditionally the problem has theoretically been addressed within the framework of continuum mechanics which relates the motion of a crack to the applied loading \cite{lawn}. For brittle solids, the problem boils down to destabilization and subsequent growth of a pre-existing crack under tension. Griffith (in 1920) \cite{griffith} considered a single sharp crack in a homogeneous elastic medium and suggested that the weakening of material by a crack could be treated as an equilibrium problem in which the reduction of strain energy, when the crack propagates, could be equated to the increase in surface energy due to the increase in surface area. He found that the critical stress $\sigma_c$ to cause a crack of length $l$, to extend is $\sigma_c = (2Yg/\pi l)^{1/2}$, where $Y$ is Young’s modulus and $g$ is the surface energy per unit area of the crack surface \cite{griffith}. The Griffith criterion has been extensively verified in engineering specimens containing cracks of controlled length and is still used to estimate the surface energy in cracks in brittle materials \cite{lawn,broberg}.  

However, the Griffith criterion is often found to lead to significant errors in the estimation of surface energies. Even for perfectly brittle materials, modifications of Griffith's theory are needed to take into account the discrete atomistic nature of the interactions \cite{broberg}. The discreteness and heterogeneities of materials lead to energy barriers at the crack tip which can arrest crack motion: a phenomenon known as lattice trapping or intrinsic crack resistance \cite{bernstein,cleri,curtin,perez,rice,thomson}. Lattice trapping in crystalline materials and intrinsic crack resistance in heterogeneous materials have received lots of attention in the context of nucleation and growth of cracks in these materials. One consequence of lattice trapping is the appearance of two length scales in the systems \cite{curtin, bernstein,cleri,long}: a large length scale associated with the elastic deformation around the crack tip and the other associated with the dissipation of energy at the crack tip. The morphology of the crack and the nature of the crack growth depends much on the length of the crack in comparison to these two length scales. A modification of the classical Griffith's theory has also been proposed \cite{wnuk} considering the discrete  nature of crack propagation which suggests: $\sigma_c \sim (2Yg/\pi(c+l))^{1/2}$, where $c$ is a measure of the average increment of the length of the crack as the external stress is increased adiabatically. A similar variation of nominal stress with the crack length has also been observed experimentally \cite{Bazant84,Bazant98,Carpinteri84,Duan03,Armstrong14} as well as numerically \cite{Nojima95}. Recently, the balance of the internal energies is explored in the mean-field limit \cite{rt22} as well as with Langevin dynamics \cite{fft22} during a crack propagation through heterogeneous systems. In our model, as we will see later, the energy balance criterion is taken care of through the interplay between local stress profile and fluctuation of strengths from point to point.  

We present here the findings from the numerical study of cracking in a pre-cracked one-dimensional fiber bundle model (FBM) \cite{hansen}. FBM is perhaps the simplest model which has been vastly used as a prototype model of fracture in heterogeneous materials. The model is guided by threshold-activated dynamics and shows different aspects of non-equilibrium statistical mechanics. The dynamics and avalanches associated with crack propagation have been explored earlier in the context of FBM \cite{drp01,pvh14,vbdb20}. In the present article, we differ from previously explored studies by inserting a crack initially inside the bundle and seeing how it affects the strength of the bundle, avalanche statistics, and spatial correlation during the failure process. The reason behind choosing one dimension is here the local stress concentration is most prominent. As we go to a higher dimension, the model approaches the mean-field limit and the presence of the pre-existing crack will be less effective. Our main findings are that we do see the signature of two length scales ($\xi$ and $\xi^{\ast}$), associated with the length of the pre-existing crack, in the context of crack propagation in the bundle. We can relate the length scales with the microscopic entities like number density and size distributions of the cracks in the bundle. Further, as long as the length $l$ of the pre-existing crack is smaller than $\xi$, the critical stress $\sigma_c$ of the bundle remains independent of $l$. For $l > \xi$, $\sigma_c \sim \sigma_0/(1+l/\xi)^{\alpha}$, where $\sigma_0$ is the critical stress of the bundle in the absence of any pre-existing crack. $\xi^{\ast}$, on the other hand, is an extreme limit of $l$ beyond which no crack, other than the pre-existing one, develops within the bundle. We show that the exponent $\alpha$ and the lengths $\xi$, $\xi^{\ast}$ depend on $\beta$ and $\gamma$, the only two parameters which characterize the fiber bundle model. $\beta$ is the strength of disorder (corresponding to the strengths of the fibers) and is the spread of the fiber strength distribution. $\gamma$ is a measure of the range up to which the stress of a broken fiber is redistributed. The onset of localization has been studied earlier in the context of fiber bundle model by tuning the disorder strength and without the presence of any pre-existing crack \cite{brr15,r21,sgh12}. In the present article, the dependence of $\xi$, $\xi^{\ast}$ and $\alpha$ on $\beta$ and $\gamma$ and the consequence of intrinsic crack resistance in crack propagation in the bundle are discussed.  

\section{Description of Fiber Bundle Model}
The fiber bundle model consists of a number of fibers (Hookean springs) held perpendicularly between two parallel bars. The bars are pulled apart with external stress $\sigma$ (force per fiber). The disorder is introduced in the model as strengths of individual fibers which are randomly assigned following a certain distribution. When the applied stress on a fiber crosses its strength threshold, the fiber breaks irreversibly. The stress of the broken fiber is then redistributed according to a certain stress redistribution scheme. After redistribution, there might be further breaking of fibers due to local enhancement of stress profile. The cascade of failure may continue leading to global failure or stops if the stresses on fibers cannot reach their threshold limit. The applied stress is to be increased to break the next weakest fiber and the process is repeated until all fibers break. 

We have considered a fiber bundle model where the strength of an individual fiber is chosen randomly from a threshold distribution. We present here the results for a power-law threshold distribution. 
\begin{equation}\label{eq1}
p(\sigma_{\rm th}) = \begin{cases}
    \displaystyle\frac{1}{2\beta\ln 10}\sigma_{\rm th}^{-1},  & (10^{-\beta} \le \sigma_{\rm th} \le 10^{\beta}) \\
    0  & ({\rm otherwise})
  \end{cases}
\end{equation}
Here $\beta$ denotes the width of the distribution or the strength of the disorder. The motivation behind choosing such a distribution is the fact that long-tailed distributions like Weibull \cite{mf99} or power-law \cite{ft28} has already been observed for the distribution of material strength. We put a crack in a one-dimensional FBM by removing $l$ consecutive fibers. If $a$ is the distance between two consecutive fibers, the crack-length due to removal of $l$ fibers will be $la$. We can set $a=1$ without losing any generality. From now on we will address $l$ as the length of the crack that is inserted in the bundle. This creates stress concentration on other unbroken fibers. The stress concentration on the fibers will depend on the scheme one is adopting for stress redistribution. We will mostly consider the local load sharing (LLS) scheme (except for the results shown in figure \ref{alpha_xi_gamma}) where the stress on the broken fiber is redistributed on the nearest neighboring unbroken fibers. 

\begin{figure}[ht]
\centering
\includegraphics[width=8cm, keepaspectratio]{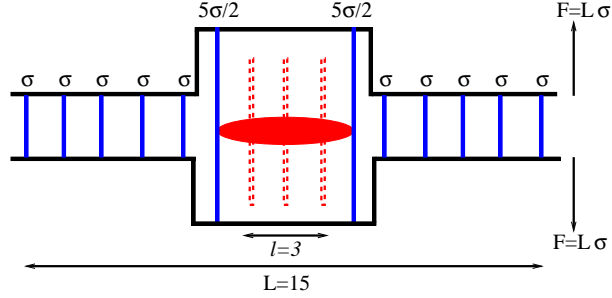} 
\caption{The figures shows the stress profile of a 1d FBM of size $L$ under local load sharing scheme (LLS) with a tensile force $F$ applied on it. A crack of length $l (=3)$ (shown by the dotted red lines) is inserted in the middle. This is equivalent to a crack of length $la$ (or $l$ if we consider $a=1$). The local stress of all fibers are $\sigma (=F/L)$ except for the two fibers at the crack tip that carries a stress $5\sigma/2$ each (see equation \ref{eq2}).}
\label{fig0}
\end{figure}

 In this situation, if the bundle is subjected to an applied stress $\sigma$, the stress on the two fibers at the crack tip due to a crack will be:
\begin{align}\label{eq2}
\sigma_{t}(l)=\sigma\left(1+\displaystyle\frac{l}{2}\right)
\end{align}
while a stress $\sigma$ is experienced by all other intact fibers. Figure \ref{fig0} shows the local stress profile for a FBM of size $L=15$ with an external force $F$ applied on it. This will create a stress $\sigma=F/L$ on the bundle in absence of any pre-existing crack. A crack of size $l=3$ is inserted in the middle by removing 3 consecutive fibers. This creates a local stress $5\sigma/2$ at the crack tips while the stress on other fibers are $\sigma$. If the stress on any fiber is higher than its threshold value, the fiber breaks irreversibly and the stress of the broken fiber is redistributed between the two nearest intact fibers at the two sides of it. For such redistribution, we have adopted an algorithm that eliminates the memory effect (similar to the algorithm in \cite{Sinha}). The algorithm we used is as follows. If a fiber, carrying stress $\sigma_b$, is broken then after redistribution the left and the right nearest neighboring unbroken fibers will experience the stresses $\sigma_l$ and $\sigma_r$ respectively as
\begin{align}\label{eq3} 
&\sigma_{l}\rightarrow \sigma_{l} + \displaystyle\frac{d_{r}}{d_{l}+d_{r}}\sigma_b \nonumber \\
&\sigma_{r}\rightarrow \sigma_{r} + \displaystyle\frac{d_{l}}{d_{l}+d_{r}}\sigma_b
\end{align}
$d_{l}$ and $d_{r}$ are the distances of the left and right neighboring intact fibers respectively from the broken fiber. 

After the redistribution, either more fibers break due to increased local stress profile or the bundle comes to a stable state where the redistributed stress cannot overcome the threshold value for any fiber. The external stress is then increased till a fiber breaks. We define $n(i) = \sigma(i)/\sigma_e$, where $\sigma(i)$ and $\sigma_e$ are the local stress on $i$th fiber and applied stress on the bundle respectively. For an external stress increment of $\Delta\sigma_e$, the local stress increment on the $i$th fiber will be    
\begin{align}\label{eq4}
\Delta\sigma(i) = n(i)\Delta\sigma_e
\end{align}
For fibers away from any broken fiber, $n(i)$ is 1. While if a fiber has an adjacent cluster of broken fibers, $n(i)$ is greater than 1 depending on the size of the cluster. The final value of $\sigma_e$ at which the global failure occurs is known as the critical stress $\sigma_c$ or strength of the bundle. 

We have also studied a more general stress redistribution scheme where the stress of a broken fiber $i$ 
is redistributed on the rest of the unbroken fibers as: 
\begin{align}\label{eq7}
\sigma(j) \rightarrow \sigma(j) + \displaystyle\frac{r_{ij}^{-\gamma}}{Z} \sigma(i)
\end{align} 
where $\sigma(j)$ is the net stress on fiber $j$ at a distance $r_{ij}$ from the broken fiber $i$. 
$Z$ is the normalizing factor and given by $Z = \displaystyle\sum_{k}r_{ik}^{-\gamma}$, where 
$k$ runs over all intact fibers. The initial local stress profile also depends on how far a certain fiber is from the assigned crack. Eq.\ref{eq2} in this case will be modified as follows:  
\begin{align}\label{eq7a}
\sigma(j) = \sigma_e \left(1 + \displaystyle\frac{R^{-\gamma}}{Z^{\prime}} l\right)
\end{align} 
Where $R$ is the distance of certain fiber from the crack. This means $R=1$ for the fibers at the notches, 2 for the next set of fibers, and so on. $Z^{\prime}$ has the form $\displaystyle\sum R^{-\gamma}$ where the sum runs over all intact fibers (which means $L-l$ fibers). In both equations \ref{eq7} and \ref{eq7a}, $\gamma$ is a measure of the extent of the stress redistribution range. A high $\gamma$ value corresponds to the LLS scheme whereas a low $\gamma$ corresponds to mean-field (MF) or equal load sharing (ELS) limit where the stress of the broken fiber is equally distributed among all other intact fibers in the bundle. A recent study \cite{brr15} shows the existence of a critical value $\gamma_c$ ($\approx 4/3$) for 1d FBM around which this change in behavior from ELS to LLS takes place. The same was observed in 2d FBM as well with a different $\gamma_c$ ($\approx 2$) \cite{hmkh02}. In the present article, in addition to the variation in $\beta$, we have also studied the crack propagation for various values of $\gamma$.
 
\section{Numerical Results} 
We present a numerical study of a pre-cracked fiber bundle model of size $10^5$ in one dimension for $10^4$ realizations and for a wide range of disorder strength $\beta$, stress relaxation range $\gamma$ and length $l$ of the pre-existing crack. We address the question: `{\it how does the size of the pre-assigned crack relate to the rupture behavior of the bundle ?}'  \\

\begin{figure*}[ht]
\centering
\includegraphics[width=6.8cm, keepaspectratio]{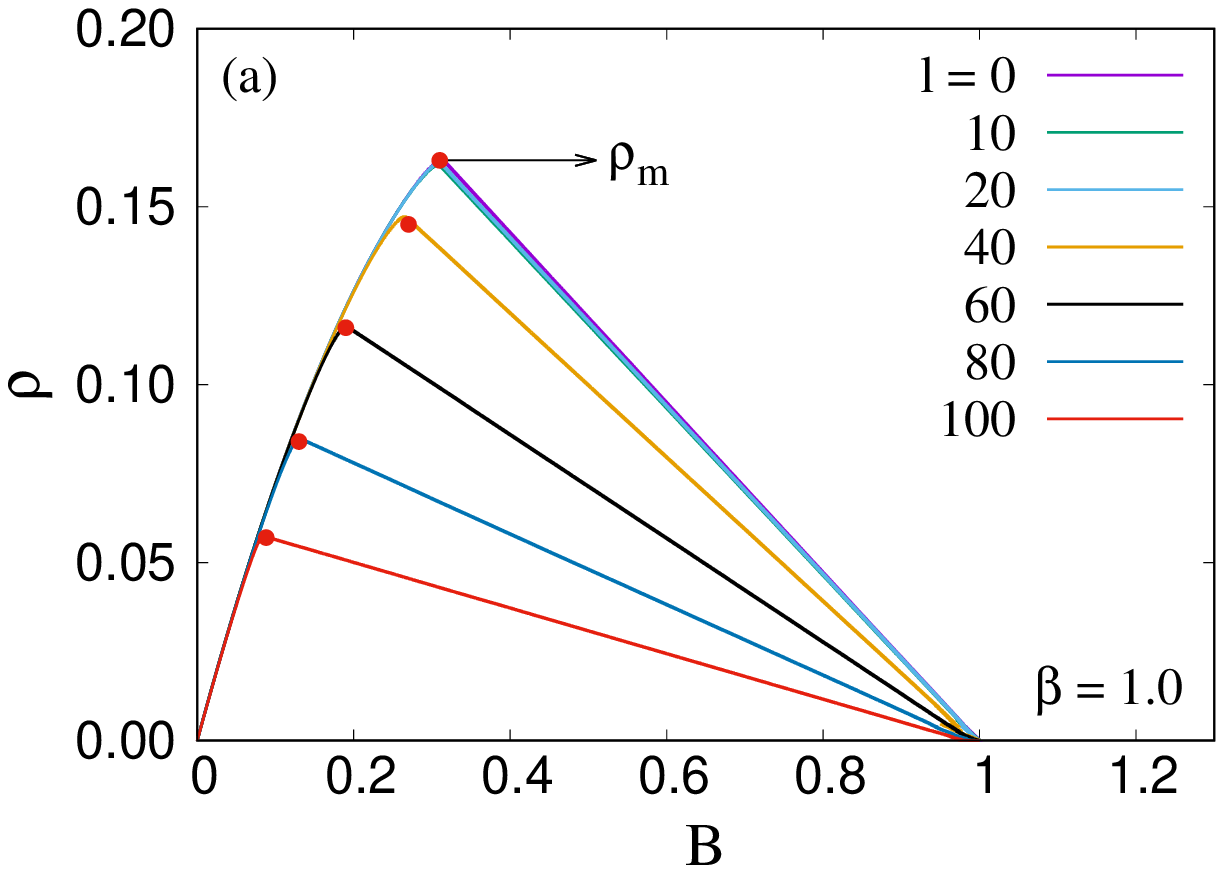} \ \ \ \includegraphics[width=6.8cm, keepaspectratio]{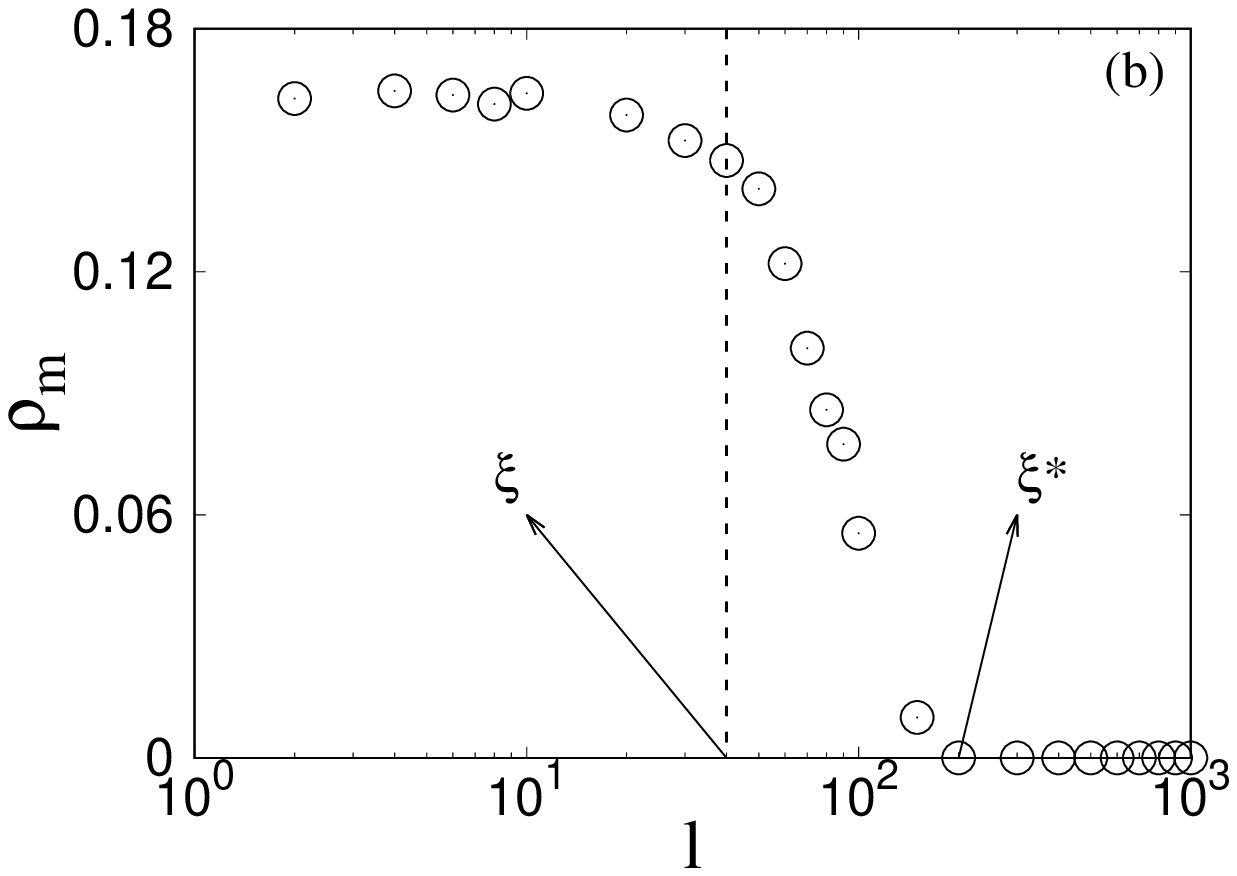} \\ \includegraphics[width=6.8cm, keepaspectratio]{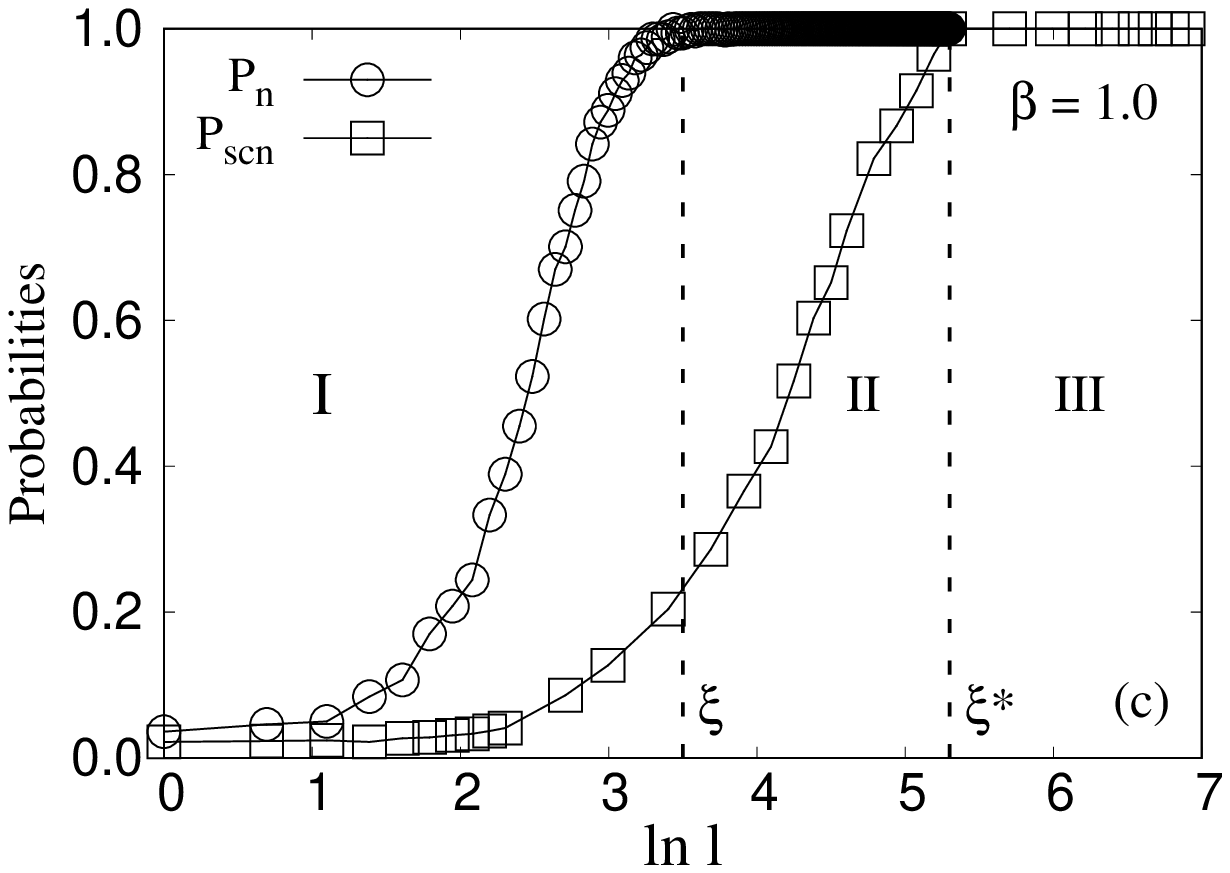} \ \ \ \includegraphics[width=6.8cm, keepaspectratio]{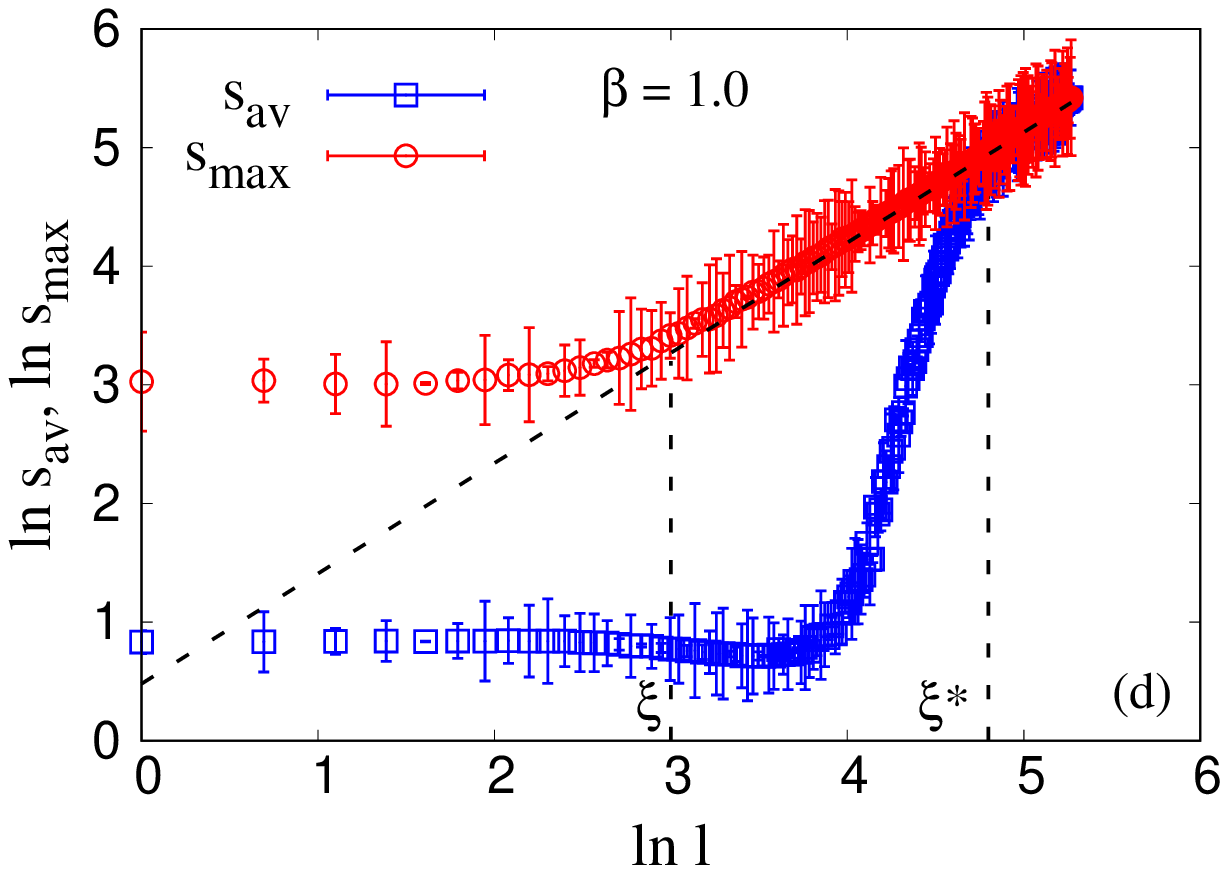} 
\caption{(a) Variation of $\rho$ with $B$ for various $l$ values and for $\beta=1.0$. The red dots stand for maximum value $\rho_m$ of the patch density. (b) Variation of $\rho_m$ with $l$ for $\beta=1.0$. For $l>\xi^{\ast}$, $\rho_m$ falls to $1/L$ (pure nucleating). For $l<\xi$, $\rho_m$ remains constant independent of l. (c) Variation of $P_n$ and $P_{scn}$ with $l$ for $\beta=1.0$. Three distinct regions I, II and III are seen. Region I: $P_{scn}$ is zero and $P_n$ is small indicating that the effect of the pre-existing crack is not observed here. Region II: $P_n$ is unity but $P_{scn}$ is low indicating that many cracks are observed here but the pre-existing crack nucleates to create global failure. Region III: Both $P_n$ and $P_{scn}$ have assumed the value unity indicating only the pre-existing crack propagates here prior to global failure. (d) Average crack size $s_{av}$ and maximum crack size $s_{max}$, just before global failure, are plotted against $l$ for $\beta=1.0$. For $l < \xi$, both $s_{av}$ and $s_{max}$ are independent of $l$. Above $l > \xi$, $s_{max} \sim l$ suggesting coalescence of the cracks to the pre-existing crack leading to fracture of the bundle. Beyond $l > \xi^{\ast}$, $s_{av}=s_{max} \sim l$, suggesting nucleation of the pre-existing crack only.}
\label{fig1}
\end{figure*}

We first present the results for the LLS fiber bundle model. In figure \ref{fig1}(a), we plot the density $\rho$ of the cracks (clusters of broken fibers) within the bundle versus the fraction of broken bonds $B$ as the bundle ruptures with time (time can be translated in this model in terms of redistributing steps). $\rho$, at a certain time $t$, is given by the ratio of the number of cracks at that time to the size $L$ of the system. $B$ is an increasing function of time as it is given by a number of broken fibers at a certain time normalized by the size $L$ of the system. $\rho$ has a value 0 and $1/L$ close to $B=0$ and $B=1$ respectively. In the former case, all fibers are intact (zero patches) and in the latter case, all fibers are broken creating a single patch. At an intermediate $B$, $\rho$ reaches a maximum value ($\rho_m$) denoted by red dots in the figure. Beyond this point, crack clusters start to coalesce and $\rho$ goes on decreasing till it reaches $1/L$ when there is a single crack in the bundle.
 
Figure \ref{fig1}(b) shows the variation of $\rho_m$ with $l$. $\rho_m$ starts at a constant value for low $l$ and decreases gradually to $\sim 1/L$ for high $l$ values. We see the signature of two length scales here. The first one is $\xi$: for $l < \xi$, $\rho_m$ is independent of $l$. This indicates that for low values of $l < \xi$, the pre-assigned crack has no effect on the fracturing of the bundle and the bundle rupture by the random breaking of the fibers in the bundle. The fracture here is dominated by the disorder and the stress concentration has no effect on the fracture process. The second length scale is $\xi^{\ast}$: for $l > \xi^{\ast}$, $\rho_m=1/L$, suggesting a pure nucleating growth where a single crack, the pre-existing one, grows rupturing the bundle. In this region, the cracking is dominated solely by stress concentration. Naturally, $\xi, \xi^{\ast}$ depend on $\beta$ which will be discussed later.

Figure \ref{fig1}(c) shows the variation of the probabilities, $P_n$ and $P_{scn}$ with $l$ for $\beta=1.0$. $P_n$ is the probability that at the final stage of fracture, the instability of the bundle is triggered by the growth of the crack which contains the pre-assigned crack. This is done by monitoring the last fiber $j$ that breaks and causes instability and calculating the probability that fiber $j$ is part of the enlarged pre-existing crack. $P_{scn}$, on the other hand, is the probability for single crack nucleation, where only one crack, the pre-existing one grows and ruptures the bundle. For low $l$ in the region I ($l<\xi$), the low value of $P_n$ tells us that the pre-assigned crack has no effect on the rupturing of the bundle. This is the region where disorder dominates the fracturing process and sets the scale $\xi$. In region III ($l > \xi^{\ast}$), $P_{scn}$ attains unity which means that statistically, only the pre-assigned crack grows in the bundle rupturing it. No other crack develops in the system. This is the extreme situation where the fracture is dominated solely by stress concentration at the two ends of the pre-assigned crack. The stress concentration sets the scale $\xi^{\ast}$. In region II, both these two scales compete. In this region, many other cracks develop in the system so that $P_{scn}$ is less than unity, but these cracks coalesce with the pre-assigned crack finally rupturing the bundle. Region II is thus dominated by both nucleation and coalescence.

Figure \ref{fig1}(d) shows how the average crack size $s_{av}$ and the maximum crack size $s_{max}$ varies with $l$ for $\beta=1.0$. The diagonal dotted line is the locus of $y \propto x$ while the vertical dotted lines show the positions of $\xi$ and $\xi^{\ast}$ respectively. For $l < \xi$, both $s_{av}$ and $s_{max}$ show no variation with $l$. $s_{max}$ is larger than $l$ and $s_{av}$ is much smaller than $l$. This suggests that there are many cracks developed in the bundle and the one that finally brings in instability and ruptures the bundle has nothing to do with the pre-existing crack. For $\xi^{\ast} > l > \xi$ , $s_{max} \sim l$ while $s_{av}$ remains much lower that $l$. This suggests that though there are large number of cracks present in the system, the largest one has size proportional to $l$. Finally, for $l > \xi^{\ast}$, $s_{av}=s_{max} \sim l$ when only the pre-existing crack grows in the bundle and no-other crack generate.\\

\begin{figure}[ht]
\centering
\includegraphics[width=6.8cm, keepaspectratio]{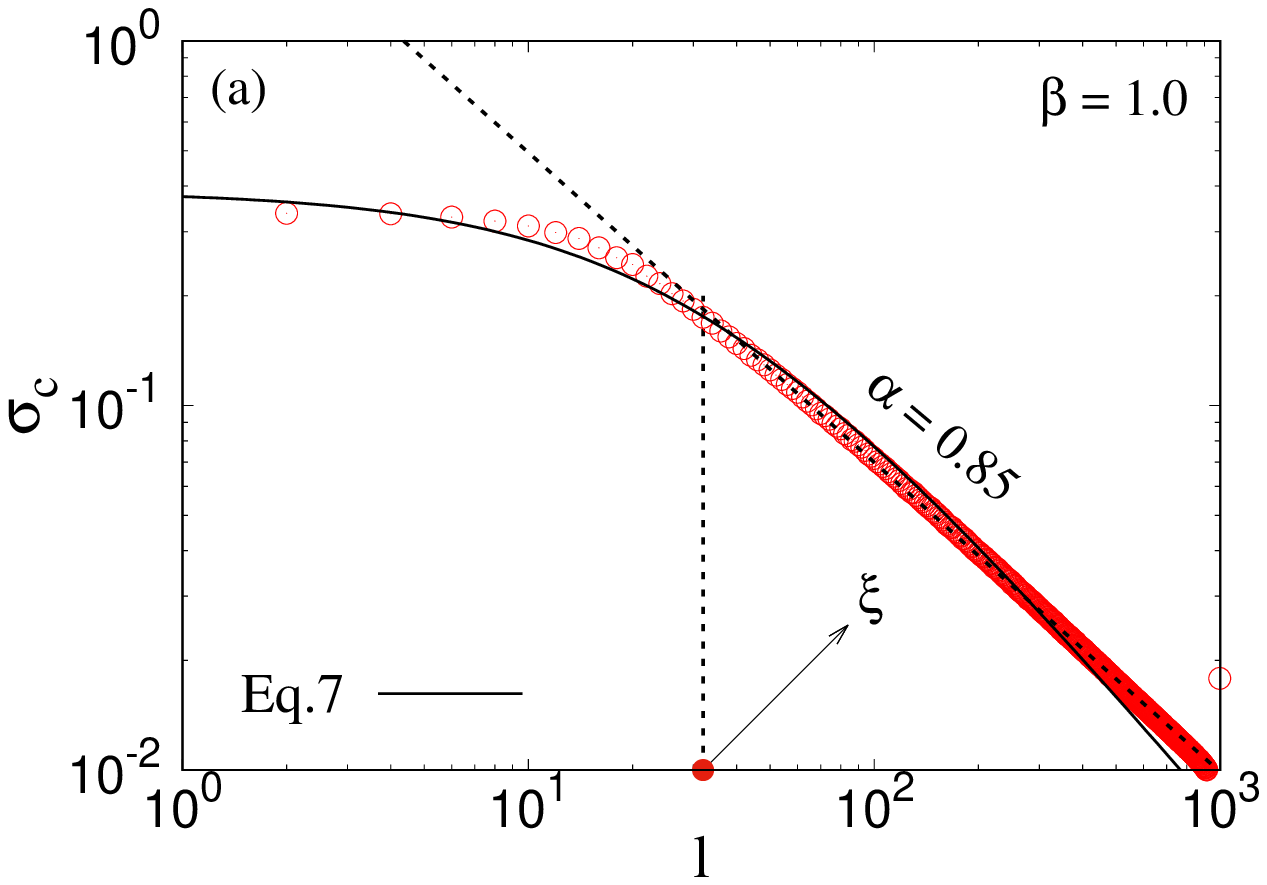} \includegraphics[width=6.8cm, keepaspectratio]{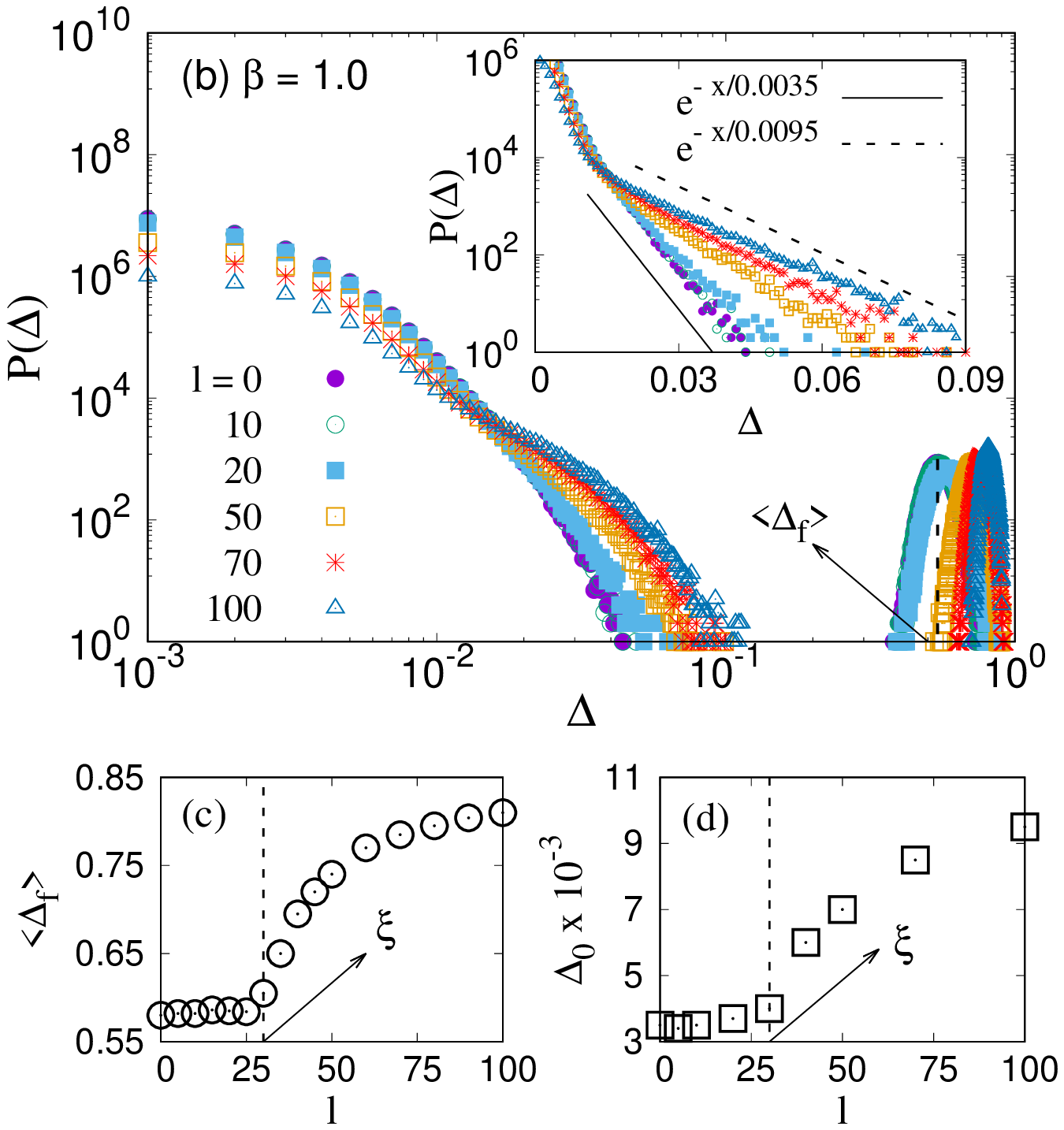}
\caption{(a) The figure shows the variation of the critical stress $\sigma_c$ with $l$ for LLS stress redistribution scheme and for $\beta=1.0$. For $l<\xi$, $\sigma_c$ is independent of $l$. For $l>\xi$, $\sigma_c \sim l^{-\alpha}$, where $\alpha \simeq 0.85$ is an exponent. (b) Avalanche size distribution $P(\Delta)$ versus $\Delta$ for different lengths $l$ of pre-existing cracks for LLS fiber bundle model at $\beta=1.0$. The inset shows the the exponential decrease of $P(\Delta)$ with $\Delta$ for small avalanche size $\Delta$ in log-normal scale. (c) The average size $\langle\Delta_f\rangle$ of the large avalanches is plotted against $l$. (d) The characteristic length $\Delta_0$ which defines the exponential fall of the distribution is plotted against $l$.} 
\label{fig4}
\end{figure}

Now that we have identified the two length scales associated with the crack development in the fiber bundle model and established links between these length scales and the microscopic entities like size and number distribution of clusters of broken fibers, we would like to discuss how the macroscopic observables (often measured in cracking experiments) behave at different length scales. To this end, we have measured the critical stress and the avalanche behavior of the bundle for various $l$ values. These two quantities have been discussed a lot in the context of crack propagation in materials \cite{brc} and in fiber bundle model \cite{hansen}. \\

Figure \ref{fig4}(a) shows the variation of critical stress $\sigma_c$ with $l$ in LLS model and for $\beta=1.0$. We find for $l < \xi$, $\sigma_c$ is independent of $l$. For $l > \xi$, $\sigma_c$ decreases with $l$ as
\begin{align}\label{eq6}
\sigma_c = \displaystyle\frac{\sigma_0}{\left(1+\displaystyle\frac{l}{\xi}\right)^{\alpha}}
\end{align} 
Here $\xi$ appears as the cut-off length and $\alpha$ is an exponent whose value depends on $\beta$ and $\gamma$. We will discuss this dependence later. As is mentioned in the introduction, this form of dependence of $\sigma_c$ on $l$ is very much seen in experiments \cite{Bazant84,Bazant98,Carpinteri84,Duan03,Armstrong14} and simulations \cite{Nojima95} and also has been suggested theoretically \cite{lawn,griffith,broberg}. It is to be noted that even for $l=0$, there is a finite value of $\sigma_c$ unlike that in Griffith's law. This is a consequence of intrinsic crack resistance.

\begin{figure}[ht]
\centering
\includegraphics[width=6.8cm, keepaspectratio]{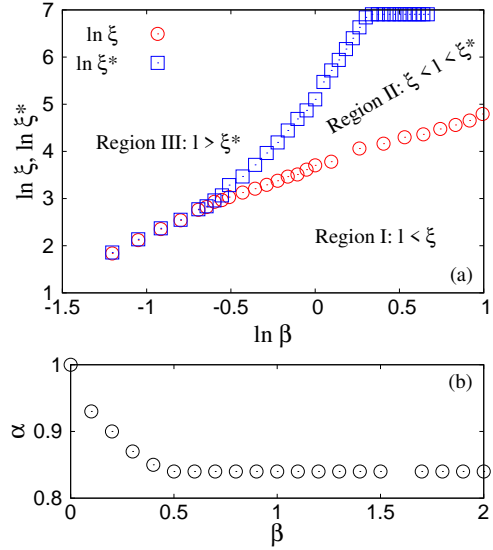} 
\caption{(a) The figure shows the variation of $\xi$ and $\xi^{\ast}$ with $\beta$ for LLS FBM. We observe three regions: I, II and III, depending on the spatial correlation during failure. Region I: conventional LLS limit where the crack does not play any role. Region II: multiple cracks but only the pre-existing crack nucleates and creates global failure. Region III: only a single crack, the pre-existing one, propagates prior to global failure. No other crack originates within the bundle.
(b) Variation of $\alpha$ with $\beta$ for LLS FBM. $\alpha$ has a high value, close to 1, for low $\beta$ and then decreases with $\beta$. For $\beta>0.5$, $\alpha$ remains constant around 0.85 independent of the disorder.
strength.}
\label{Delta2_vs_Beta}
\end{figure}

Figure \ref{fig4}(b) shows the avalanche size distribution for different lengths $l$ of the pre-assigned crack in LLS fiber bundle at $\beta=1.0$. An avalanche is defined as the number of fibers broken in between two consecutive stress increments. The avalanches in the fiber bundle model show a robust behavior in its size distribution \cite{hh92,khh97} as well as have information about upcoming catastrophic events \cite{kpk20}. We have presented the results in terms of $\Delta$, which is the avalanche size or the number of broken fibers in one avalanche normalized by size $L$ of the system. Figure \ref{fig4}(b) shows the variation of $P(\Delta)$ with $\Delta$. In abrupt brittle failure at low disorder (both ELS \cite{rr15} and LLS \cite{r17} FBM), there is no need for stress increment after the weakest fiber is broken. In this limit, the bundle breaks in a single avalanche. On the other hand, at a moderate disorder, the avalanche size distribution shows a scale-free behavior with a universal exponent -5/2 \cite{hh92} and an exponential nature \cite{khh97} for ELS and LLS scheme respectively. We observe the same exponential decay of the avalanche size distribution $P(\Delta) \sim \exp(-\Delta/\Delta_0)$ for finite-size avalanches and distribution for largest avalanche $\Delta_f$ prior to rupture even if there is a pre-existing crack within the bundle. In the regime, $l<\xi$, the characteristic length $\Delta_0$ for the exponential decay and the average value $\langle\Delta_f\rangle$ of the largest avalanche remain almost constant. However, both these quantities increase rapidly with $l$ for $l > \xi$ (see figure \ref{fig4}(c) \& (d)). For large $l$, the rapid exponential fall of finite avalanches and a large value of the large avalanche indicates that the bundle ruptures by destabilization of the crack by large avalanches. \\

\begin{figure}[ht]
\centering
\includegraphics[width=6.8cm, keepaspectratio]{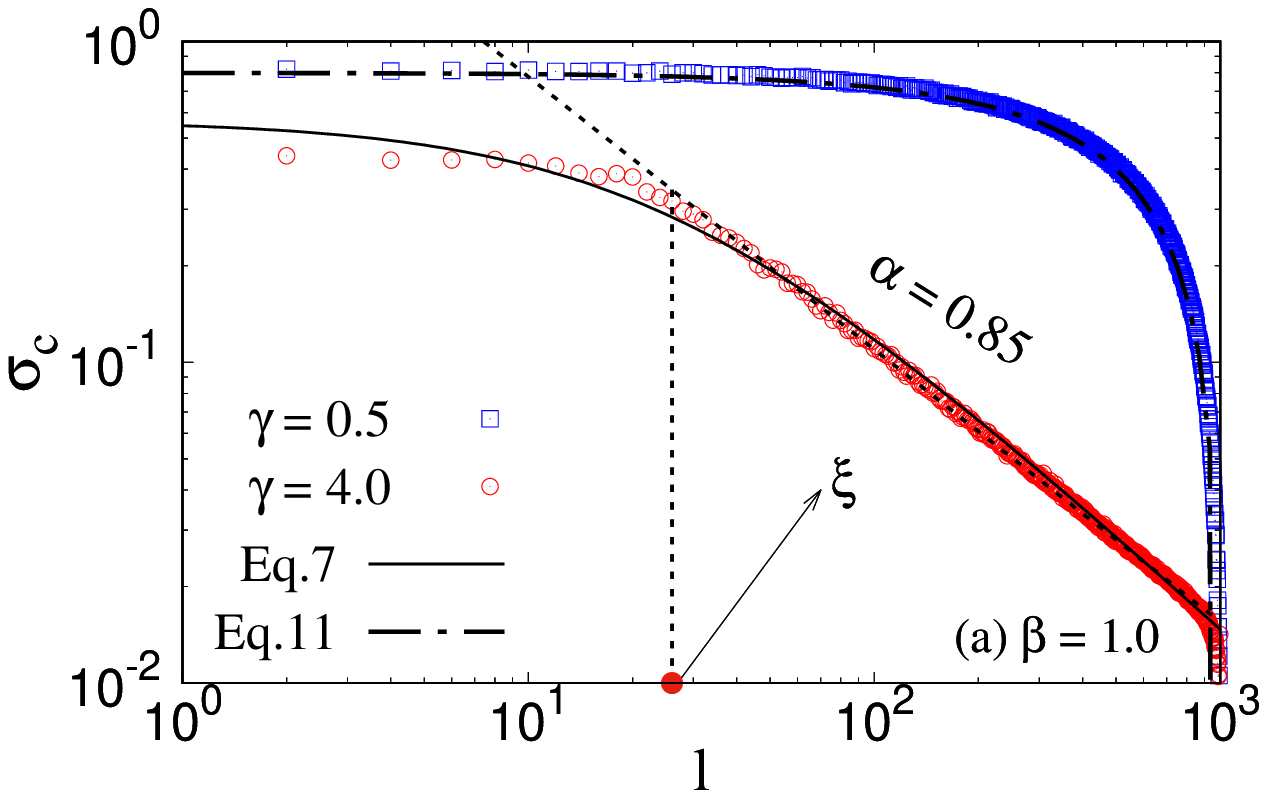} \includegraphics[width=6.8cm, keepaspectratio]{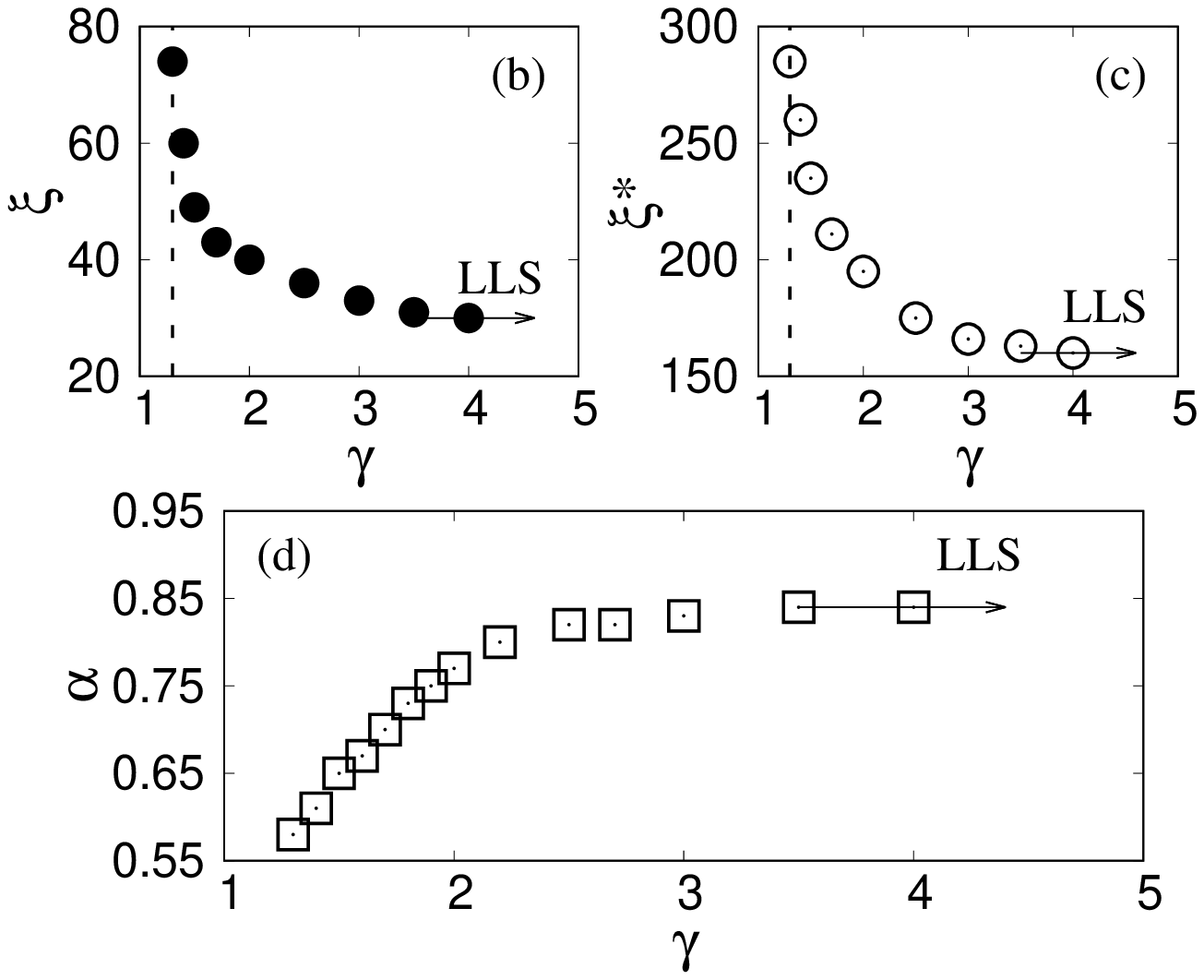} 
\caption{(a) Variation of $\sigma_c$ with $l$ for $\gamma=0.5$ ($<\gamma_c$) and $\gamma=4.0$ ($>\gamma_c$) and for $\beta=1.0$. The behavior matches with Eq.\ref{eq6} (LLS FBM) and Eq.\ref{eq11} (MF FBM) when $\gamma$ is 4.0 and 0.5 respectively. (b), (c) and (d) respectively shows the variation of $\xi$, $\xi^{\ast}$ and $\alpha$ with $\gamma$. High $\gamma$ corresponds to the LLS limit. On the other hand, when $\gamma$ is low, both $\xi$ and $\xi^{\ast}$ increases with decreasing $\gamma$. This is due to the fact that in low $\gamma$, the local stress concentration does not play a crucial role and the pre-existing crack will be visible to the model only when the length of the crack is large. $\alpha$ decreases with decreasing $\gamma$ suggesting less effect of the crack on the critical stress.}
\label{alpha_xi_gamma}
\end{figure}

We would now discuss how the length scales and the exponent $\alpha$ depend on $\beta$. Figure \ref{Delta2_vs_Beta}(a) shows the variation of above mentioned two characteristic lengths, $\xi$ and $\xi^{\ast}$, when the strength of disorder $\beta$ is increased. We observe three distinct regions. Region I: $l<\xi$ and the pre-assigned crack has no effect on the cracking of the bundle. The critical stress is independent (see figure \ref{fig4}) of the crack length. The dynamics of the system in absence of any pre-assigned crack has been explored earlier in detail \cite{rbr17}. This region is dominated by the disorder and stress redistribution has not much effect on cracking. For high $\beta$, the rupture sequence of the fibers is random, whereas, for low $\beta$, cracks nucleate at suitable soft spots independent of the pre-assigned crack. Region II: In this region, disorder competes with stress redistribution in cracking. Failure is dominated by crack length. The critical stress falls with $l$ in a scale-free manner with an exponent $\alpha$ (see figure \ref{fig4}). The probability that the pre-existing crack causes final instability in the bundle is unity (see figure \ref{fig1}c), which means the pre-existing crack nucleates though there are other cracks developed within the bundle. Region III: This region is dominated by stress redistribution. This is an extreme limit of the crack-length dominated region. Here only the pre-existing crack propagates within the bundle and no other crack is originated. The critical stress follows the scale-free decay with $l$ here as well. One should notice that for low $\beta$, $\xi$ and $\xi^{\ast}$ coincide with each other. In this limit, the disorder strength is not strong enough to trap the pre-existing crack unless the length of the crack is small itself. As a result, we either see random failure events (region I) at low $l$ or pure nucleating failure beyond a certain length scale (region III). The increase in length scales with $\beta$ makes sense as a large crack has to be inserted within the bundle at high disorder so that the stress concentration overcomes the high fluctuation in threshold strength values, in other words, the intrinsic crack resistance.    

Figure \ref{Delta2_vs_Beta}(b) shows the variation of $\alpha$ with $\beta$ for LLS FBM. $\alpha$ has a high value 1 for $\beta=0$, which is consistent with equation \ref{eq2} since all fibers have the same threshold and the entire bundle ruptures in a single avalanche as soon as the weakest fiber is broken. As $\beta$ increases, $\alpha$ starts to decrease and remains constant around 0.85 independent of the disorder beyond $\beta=0.5$. \\

So far, we have presented our findings for a particular value of $\beta$ and in the LLS fiber bundle model. In principle, we 
have two parameters in the model: the disorder measured by its strength $\beta$ and the range of stress redistribution characterized by $\gamma$. We have seen that the origin of two length scales is essentially due to the competition between 
stress concentration on the fibers due to stress redistribution in the cracking process and the disorder in the threshold strengths of the fibers. We would expect that our results will also depend on the stress redistribution range characterized by $\gamma$ in addition to the disorder strength characterized by the width of the threshold distribution of the fibers. 

We now present our results for the variation of $\xi$, $\xi^{\ast}$ and $\alpha$ with $\gamma$. We have already discussed that a low value of $\gamma$ corresponds to the MF or ELS version of the model. The critical stress, in presence of a pre-existing crack, can be calculated in the MF limit analytically. We will then compare the analytical result with the numerical behavior of critical stress with $l$ keeping $\gamma=0.5$ (which is much below $\gamma_c=4/3$ for 1d FBM \cite{brr15}). For fiber strength distribution as given by \eqref{eq1}, the cumulative distribution is given by
\begin{align}\label{eq8}
P(\sigma_{th}) = \displaystyle\int_{10^{-\beta}}^{\sigma_{th}} p(\sigma_{th}) d\sigma_{th} = \displaystyle\frac{\ln \sigma_{th}}{2\beta \ln 10} + \displaystyle\frac{1}{2}
\end{align}    
where $p(\sigma_{th})=\sigma_{th}^{-1}/(2\beta\ln 10)$ is the normalized density function. For applied force $F$ and an extension $\epsilon$ of the fibers, functional form between $F$ and $\epsilon$ can be expressed as  
\begin{align}\label{eq9}
F(\epsilon) = L\epsilon\left(\displaystyle\frac{1}{2}+\displaystyle\frac{\ln \epsilon}{2\beta\ln 10}\right)
\end{align} 
The function $F(\epsilon)$ has a maximum at critical extension $\epsilon_c=10^{\beta}/e$ ($e \approx 2.72$), satisfying 
$dF(\epsilon)/d\epsilon=0$. The critical value $\sigma_t(l)$ required to break the model in presence of the crack of length $l$ is generated by a critical external stress $\sigma_c$, 
\begin{align}\label{eq10}
\sigma_t(l) = \sigma_c\left(1+\displaystyle\frac{l}{L-l}\right) = \displaystyle\frac{F_c}{L} = \displaystyle\frac{F(\epsilon_c)}{L} = \displaystyle\frac{10^{\beta}}{2\beta e \ln 10}
\end{align}
The critical stress $\sigma_c$ is given as follows
\begin{align}\label{eq11}
\sigma_c = \displaystyle\frac{10^{\beta}}{2\beta e \ln 10}\left(1-\displaystyle\frac{l}{L}\right)
\end{align}
This shows that the critical stress increases with disorder strength $\beta$ and decreases with crack-length $l$. \\

Figure \ref{alpha_xi_gamma} shows how the cracking behavior depends on the stress release range $\gamma$. Figure \ref{alpha_xi_gamma}(a) shows for a high value of $\gamma$ (= 4.0), $\sigma_c$ shows the same behavior with $l$ as in the LLS fiber bundle model: $\sigma_c$ is independent of $l$ for  $l< \xi$ and decays with a power law for $l > \xi$. The behavior for $\sigma_c$ at high $\gamma$ is fitted with Eq.\ref{eq6} with $\xi=28$ and $\alpha=0.85$. Both $\xi$ and $\alpha$ are functions of $\gamma$. For a low $\gamma$ (= 0.5) value, the model is in the MF (or ELS) limit and we observe $\sigma_c$ versus $l$ behavior to be very close to the mean-field expression given by Eq.\ref{eq11}.

Figure \ref{alpha_xi_gamma}(b), (c) and (d) show the variations of $\xi$, $\xi^{\ast}$ and $\alpha$ respectively with $\gamma$. In the limit of high $\gamma$, all the three parameters $\xi$, $\xi^{\ast}$ and $\alpha$ saturates towards the values we obtained for the LLS fiber bundle model. As $\gamma$ decreases both $\xi$ and $\xi^{\ast}$ increases. This happens as with decreasing $\gamma$ the effect of stress concentration at the notches of the pre-existing crack becomes less effective and a larger crack is required to make the stress concentration effective again. This in turn increases the sizes of the characteristic length scales which finally diverges around $\gamma_c$ (the dotted lines in figure \ref{alpha_xi_gamma}(b) and (c)). $\alpha$ on the other hand decreases as $\gamma$ decreases up to $\gamma \ge \gamma_c$. This is also due to the fact that the stress concentration becomes less effective as $\gamma$ increases and that increases the chances of more resistance to cracking. This lowers the value of $\alpha$. Due to the same reason, earlier $\alpha$ was observed to decrease with an increase in $\beta$ as well. Below $\gamma_c$, as the model is in the mean-field limit, the behavior of $\sigma_c$ is not scale-free and $\alpha$ does not have meaning there. \\    

\begin{figure}[ht]
\centering
\includegraphics[width=6.8cm, keepaspectratio]{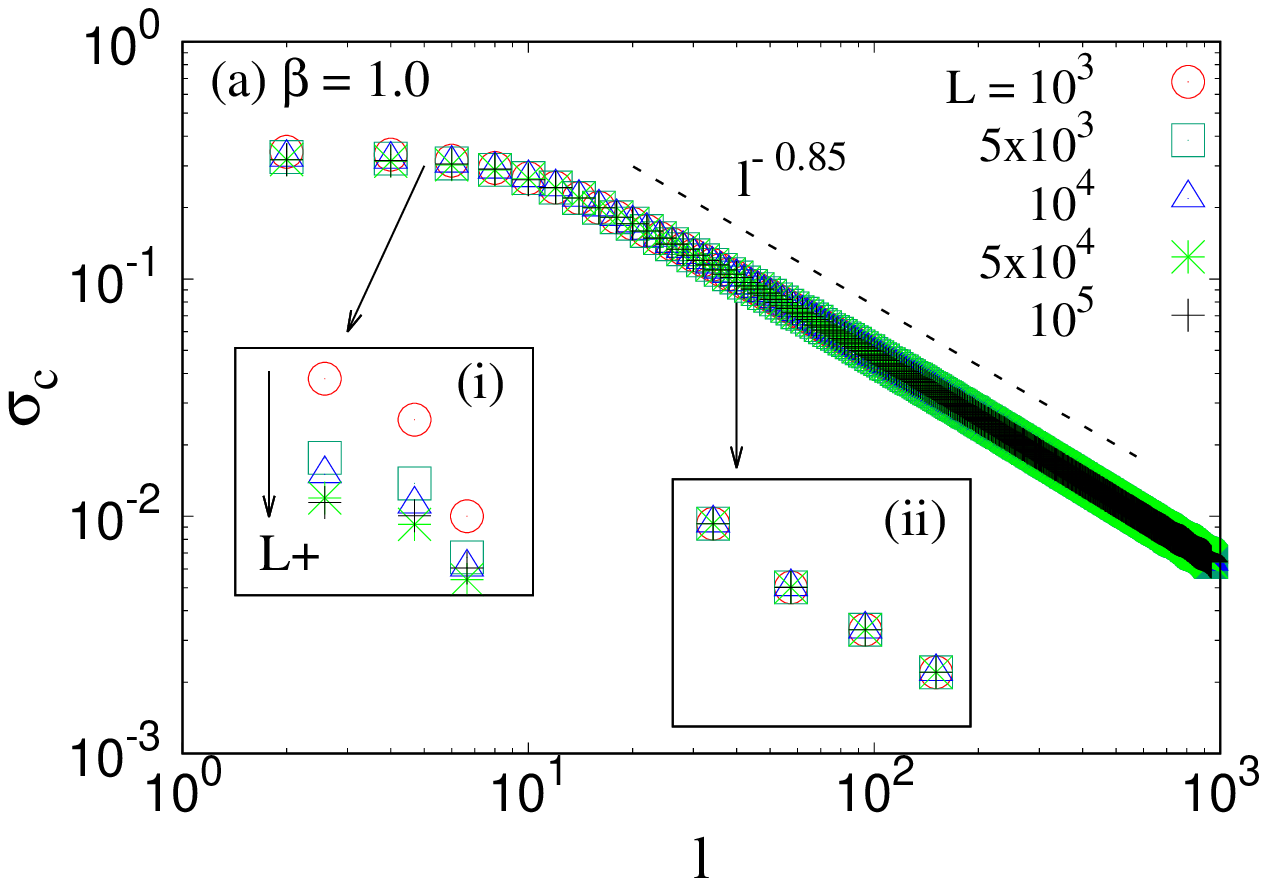}  \includegraphics[width=6.8cm, keepaspectratio]{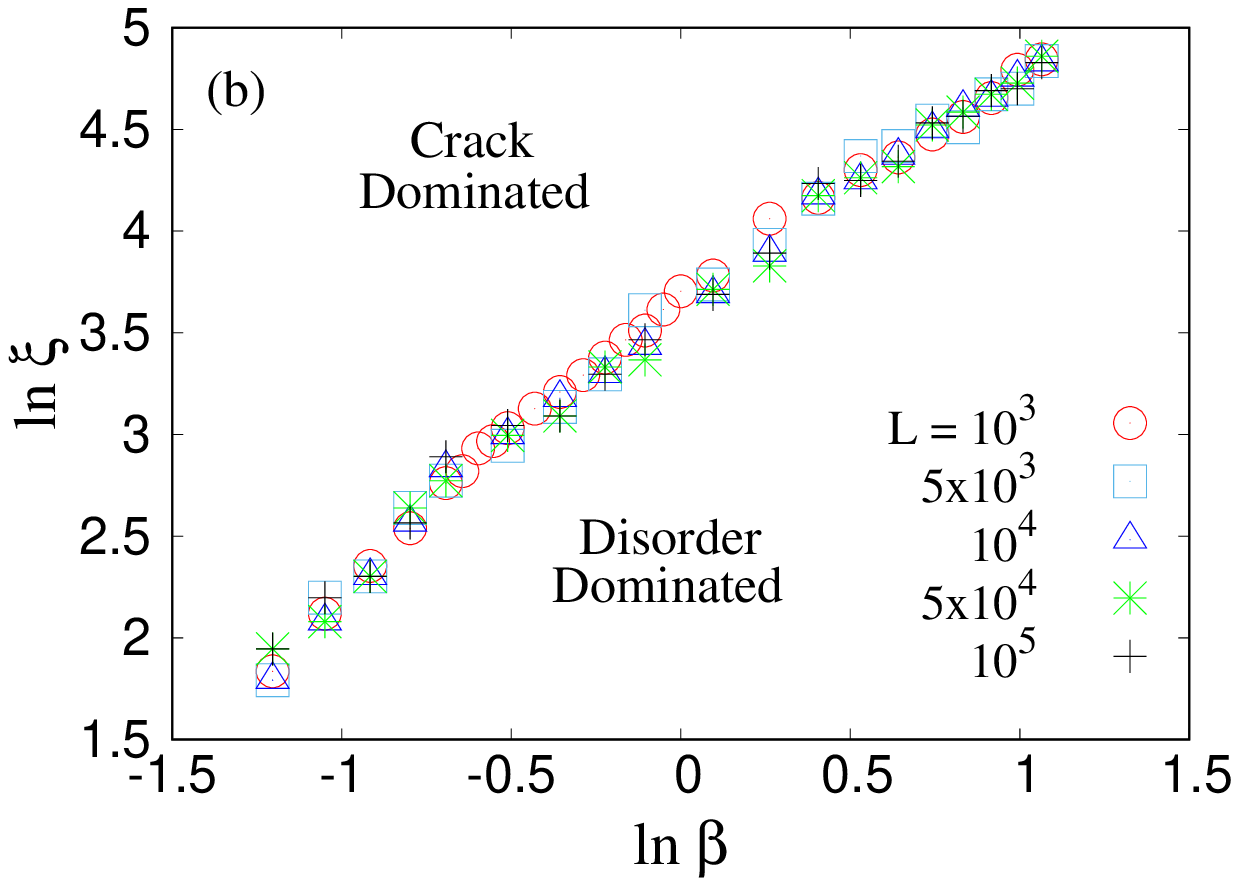}
\caption{(a) $\sigma_c$ vs $l$ for system sizes ranging in between $10^3$ and $10^5$. For $l<\xi$, $\sigma_c$ is independent of $l$ but decreases with increasing $L$ (see inset (i)). The system size effect is not visible in the region $l>\xi$ (see inset (ii)). (b) Variation of characteristic length $\xi$ with disorder strength $\beta$ for system sizes ranging in between $10^3$ and $10^5$. $\xi$ is not observed to change with size of the system. We set $\beta=1.0$.}
\label{fig10}
\end{figure}

We briefly discuss any possible effect of system size on our results. Figure \ref{fig10}(a) shows the variation of $\sigma_c$ with $l$ for system sizes $10^3$, $5 \times 10^3$, $10^4$, $5 \times 10^4$ and $10^5$ with LLS scheme. $\beta$ is kept constant at 1.0. When $l<\xi$, a reduction in critical stress is observed with the increasing size of the bundle (see the inset (i) of the same figure). This reduction in $\sigma_c$ is observed due to the {\it weakest link of chain} like failure dynamics in this model as the weakest link itself scales to lower values as the size of the bundle increases \cite{hansen}. On the other hand, we do not see any system size dependence $\xi$ for $l > \xi$ (see the inset (ii) of the same figure). The probable reason for this is in this limit the critical stress is already lower than the value we observe due to the weakest link effect. This happens due to the fact that a very small external stress can create large stress at the crack notches. The redistributed stress depends on the crack-length and disorder strength and not on the system size. Though we don't claim the same size-effect of $\sigma_c$ at higher dimensions since the system size scaling rule of critical stress is not valid there \cite{Sinha}. Figure \ref{fig10}(b) explicitly shows that the size of the system does not have any effect on this characteristic length $\xi$. Both the regions, {\it disorder-dominated} and {\it crack-dominated}, remain unchanged when we increase the system size and we can conclude both these regions will survive in the thermodynamic limit as well. \\

The universal behavior of our results has been explored briefly for uniform and Weibull distribution for fiber strengths. The preliminary study does not show any dependence of the results on the nature of the distribution. However, this aspect needs to be studied in detail.

\section{Discussion}

In conclusion, we have mimicked the strength experiments generally done by stressing a specimen with a pre-existing crack by a numerical study of stressing fiber bundle model with a pre-existing crack in it. The important observations from cracking in engineering samples are the pronounced effect of lattice trapping or crack resistance, modification of Griffith's law, and occurrence of two length scales arising from the competition between the large scale elastic stress relaxation and short scale energy dissipation near the crack tip. The fiber bundle model is perhaps the simplest model for fracture in heterogeneous materials having essentially two parameters: the disorder in the fiber strength thresholds and regime over which the stress of a broken fiber is redistributed. We clearly see the existence of two length scales associated with the cracking of the bundle. We relate the two length scales with the microscopic observables like the number and size distribution of the cracks in the bundle. We see the modification of Griffith's law similar to what is found in engineering specimens and we determine the range of validity of the law in the case of the fiber bundle model. The other spectrum of this is the high disorder scenario where the failure events are random in space. This has been observed numerically in 2d random register network \cite{mohaha12,szs13} as well as in random spring network \cite{rd96}. Finally, we have determined the dependence of these two length scales and the exponent that characterizes the modified Griffith's law on the strength of the disorder and the range of stress redistribution from a failed fiber in the bundle. 

\section{Acknowledgements}

SR was supported by the Research Council of Norway through its Centres of Excellence funding scheme, project number 262644. TH was supported by Japan Society for the Promotion of Science (JSPS) Grants-in-Aid for Scientific Research (KAKENHI), Grants Nos. 16H06478 and 19H01811. PR acknowledges JSPS Invitational Fellowships for Research in Japan and Earthquake Research Institute, Tokyo for infrastructural support for the visit.


\begin{thebibliography}{99}
\bibitem{lawn} Lawn, B. R., \& Wilshaw T. R. Fracture of Brittle Solids. Cambridge University Press, Cambridge (1975). 
\bibitem{griffith} Knott, J. F. Fundamentals of Fracture Mechanics. Butterworths (1973). 
\bibitem{broberg} Broberg, K. B. Cracks and Fracture. Academic Press, New York (1999). 
\bibitem{bernstein} N. Bernsteina, and D. W. Hess, Phys Rev Lett {\bf 91}, 025501 (2003).
\bibitem{cleri} A. Mattoni, L. Colombo, and F. Cleri, Phys Rev Lett {\bf 95}, 115501 (2005).
\bibitem{curtin} W. A. Curtin, J Mater Res {\bf 5}, 1549 (2000). 
\bibitem{perez} R. Perez and P. Gumbsch, Phys Rev Lett {\bf 84}, 5347 (2000).
\bibitem{rice} J. R. Rice, J Mech Phys Solids {\bf 40}, 239 (1992).
\bibitem{thomson} R. Thomson, C. Hsieh, and V. Rana, J Appl Phys {\bf 42}, 3154 (1971).
\bibitem{long} R. Long, C. Y. Hui, J. P. Gong, and E. Bouchbinder, Ann. Rev. of Cond. Matt. Phys. {\bf 12}, 71 (2021).
\bibitem{wnuk} M. P. Wnuk, and A. Yavari, Eng. Fract. Mech. {\bf 75}, 1127 (2008).
\bibitem{Bazant84} Z. P. Bazant, Metal Journal of Eng. Mec., Vol. 110, 518-535 (1984). 
\bibitem{Bazant98} Z. P. Bazant and J. Planas, Fracture and size effect in concrete and other quasibrittle materials. CRCPress (1997).
\bibitem{Carpinteri84} A. Carpinteri, Theo. and App. Frac. Mech. {\bf 2} 39 (1984). 
\bibitem{Duan03} K. Duan, X. Z. Hu and F. H. Wittmann, Materials and Structures/Materiaux et Constructions {\bf 36}, 74 (2003).
\bibitem{Armstrong14} R. W. Armstrong, Phil. Trans. R. Soc. A {\bf 373}, 20140124 (2014).
\bibitem{Nojima95} T. Nojima, and H. Tsuyoshi, J. Soc. Mat. Sci., Japan {\bf 44}(499), 451-456 (1995).
\bibitem{rt22} H. B. da Rochaa, and L. Truskinovsky, J. Mec. Phys. of Sol. 158, 104646 (2022).
\bibitem{fft22} Y. Feng, J. Fan, and E. B. Tadmor, J. Mec. Phys. of Sol. 159, 104715 (2022).
\bibitem{hansen} A. Hansen, P. C. Hemmer, and S. Pradhan, The Fibre Bundle Model. Weinheim, Germany: Wiley-VCH (2015).
\bibitem{drp01} A. Delaplace, S. Roux, G. Pijaudier-Cabot, Journal of engineering mechanics, {\bf 127(7)}, 646 (2001).
\bibitem{pvh14} S. Patinet, D. Vandembroucq, and A. Hansen, Eur. Phys. J. Spec. Top. {\bf 223}, 2339 (2014).
\bibitem{vbdb20} F. Villette, J. Baroth, F. Dufour, J. F. Bloch, and S. R. Du Roscoat, International Journal of Fracture, {\bf 221(1)}, 87 (2020). 
\bibitem{brr15} S. Biswas, S. Roy, and P. Ray, Phys Rev E {\bf 91}, 050105(R) (2015).
\bibitem{r21} S. Roy, Front. Phys. {\bf 9}, 752086 (2021).
\bibitem{sgh12} A. Stormo, K. S. Gjerden, and A. Hansen, Phys. Rev. E {\bf 86}, 025101(R) (2012). 
\bibitem{mf99} D. Munz and T. Fett, Ceramics: Mechanical Properties, Failure Behavior, Materials Selection. Springer-Verlag Berlin Heidelberg (1999).
\bibitem{ft28} R. A. Fisher, and L. H. C. Tippett, Proc. Cambridge Philos. Soc. {\bf 24}: 180–190 (1928).
\bibitem{Sinha} S. Sinha, J. T. Kjellstadli and A. Hansen, Phys Rev E {\bf 92}, 020401(R) (2015).
\bibitem{hmkh02} R. C. Hidalgo, Y. Moreno, F. Kun, and H. J. Herrmann, Phys. Rev. E {\bf 65}, 046148 (2002).
\bibitem{brc} S. Biswas, P. Ray, and B. K. Chakrabarti, Statistical Physics of Fracture Breakdown and Earthquake, Wiley-VCH, Berlin Germany (2015).
\bibitem{hh92} P. C. Hammer and A. Hansen, ASME J Appl Mech {\bf 59}, 909 (1992).
\bibitem{khh97} M. Kloster, A. Hansen, and P. C. Hemmer, Phys Rev E {\bf 56}, 2615 (1997).
\bibitem{kpk20} V. Kadar, G. Pal, and F. Kun, Scientific Reports {\bf 10}, 2508 (2020).
\bibitem{rr15} S. Roy, and P. Ray, Europhysics Letters {\bf 112}, Number 2, 26004 (2015).
\bibitem{r17} S. Roy, Phys Rev E {\bf 96}, 042142 (2017).
\bibitem{rbr17} S. Roy, S. Biswas, and P. Ray, Phys Rev E {\bf 96}, 063003 (2017). 
\bibitem{mohaha12} A. A. Moreira, C. L. N. Oliveira, A. Hansen, N. A. M. Araujo, H. J. Herrmann, and J. S. Andrade Jr., Phys. Rev. Lett. {\bf 109}, 255701 (2012).
\bibitem{szs13} A. Shekhawat, S. Zapperi, and J. P. Sethna, Phys. Rev. Lett. {\bf 110}, 185505 (2013).
\bibitem{rd96} P. Ray and G. Date, Physica A: Statistical Mechanics and its Applications {\bf 229}, 26 (1996).
\end{thebibliography}
\end{document}